\documentclass[11pt,twoside]{article}

\usepackage{asp2014}

\aspSuppressVolSlug
\resetcounters

\begin{document}

\title{Extragalactic stellar photometry and the blending problem}

\author{Carlos~Feinstein,$^{1,2}$, Gustavo~Baume$^{1,2}$, Jimena~Rodr\'{\i}guez$^{1,2}$, and Marcela~Vergne$^{1,2}$
  \affil{$^1$Facultad de Ciencias Astron\'omicas y Geof\'{\i}sicas, Observatorio Astr\'onomico, UNLP, La Plata, Argentina; \email{cfeinstein@fcaglp.unlp.edu.ar}}
  \affil{$^2$Instituto de Astrof\'{\i}sica de La Plata - CONICET}}           

\paperauthor{Carlos~Feinstein}{cfeinstein@fcaglp.unlp.edu.ar}{0000-0003-2341-0494}{Observatorio Astron\'omico}{Facultad de Ciencias Astron\'omicas y Geof\'{\i}sicas, UNLP - IALP, Conicet}{La Plata}{}{1900}{Argentina}

\paperauthor{Gustavo~Baume}{gbaume@fcaglp.unlp.edu.ar}{0000-0002-0114-0502}{Observatorio Astron\'omico}{Facultad de Ciencias Astron\'omicas y Geof\'{\i}sicas, UNLP - IALP, Conicet}{La Plata}{}{1900}{Argentina}

\paperauthor{Jimena~Rodr\'{\i}guez}{jimerodriguez@fcaglp.unlp.edu.ar}{}{Observatorio Astron\'omico}{Facultad de Ciencias Astron\'omicas y Geof\'{\i}sicas, UNLP - IALP, Conicet}{La Plata}{}{1900}{Argentina}

\paperauthor{Marcela~Vergne}{mvergne@fcaglp.unlp.edu.ar}{}{Observatorio Astron\'omico}{Facultad de Ciencias Astron\'omicas y Geof\'{\i}sicas, UNLP - IALP, Conicet}{La Plata}{}{1900}{Argentina}

  
\begin{abstract}
			The images provided by the Advanced Camera for Surveys at the Hubble Space Telescope (ACS/HST) have the amazing spacial resolution of 0".05/pixel. Therefore, it is possible to resolve individual stars in nearby galaxies and, in particular, young blue stars in associations and open clusters of the recent starburst. These data are useful for studies of the extragalactic young  population  using color magnitude diagrams (CMD) of the stellar groups. However, even with the excellent indicated spatial resolution, the blending of several stars in crowded fields can change the shape of the CMDs. Some of the blendings could be handled in the cases they produce  particular features on the stellar PSF profile (e.g. abnormal sharpness, roundness, etc). But in some cases, the blend could be difficult to detect, this is the case, were a pair or several stars are in the same line of sight (e.g. observed in the same pixel).      
In this work, we investigated the importance of the blending effect in several crowded regions, using both numerical simulations and real ACS/HST data. In particular, we evaluated the influence of this effect over the CMDs, luminosity functions (LFs) and reddening estimations obtained from the observations. 
  
\end{abstract}

\section{Introduction and observations}
As new high spatial resolution data have been obtained from nearby galaxies, some new problems have to be taken in account at the instance of analyzing the data. One of them is the blending, and more the important, the blending affects on the observation parameters of the PSF (e.g two stars in same sight of view and without being detected). Some of the blending can be easily found analyzing sharpness, roundness, crowding parameters which are in the typical output of the photometry software. But, we want to focus in the case that the two stars are centered in  the same pixel. So, is the undetected blending  a real a problem?. By modeling it with data (density, stars colors, etc) from real observations of the  ACS/HST we can check how probable is this happening in real data, how it would affect the observations and the results, for example, the CMD of the young clusters observed in nearby galaxies.

Data used in this work were obtained from the ACS/HST. They were collected and reduced by the ACS Nearby Galaxy Survey (ANGST - PI:Dalcanton). The selected data were download from the STSCI/ANGST site and they included photometric information for several fields covering the galaxies NGC 247, NGC 253, NGC 300 and NGC 2366. The ANGST sample was defined in \citep{2009ApJS..183...67D}.

\section{Probability of a blend}
One way to estimate the number of blends is to make Monte Carlos simulations of an stellar population over a virtual CCD with spatial resolution and area equal to the real CCD on the ACS/HST, simulating real observations. Random stars are located using a uniform probability distribution, that has the maximum stellar density observed in the real ACS/HST Images. This maximum stellar density (see table 1) was obtained observing the real data from the galaxies to a certain magnitude.
So, the blend of the sample can be easily computed in any simulations as the number of stars is chosen to reproduce the maximum density.
In the real case, this number of blends would be valid at the place in the galaxy with the maximum stellar density and a maximum level for all regions with lower density (see for example, \citep{Kiss}).  
But the same result can be more easily calculated because the uniform random distributions follow a Poisson spatial distribution over each pixel. Therefore
 its characteristic parameter is the stellar surface density ($\rho$) and the probability of having any star or one without a blend in a pixel is trivial to calculate. If $N$ is the number of blends, the Possionian probability for a blend o more is  $P(N>1)= 1-e^{-\rho} (1+\rho)$.

So, stellar density is the key parameter, that would make that the stars in nearby galaxies are resolved individuality or blended. \textbf{Table 1} shows the measured density in some galaxies of interest. The maximum density refers to the high value measured over the galaxy and its correspond to the crowded central regions. On the other hand, average density correspond to the disc typical density on each galaxy.

	\articlefigure[width=0.8\textwidth]{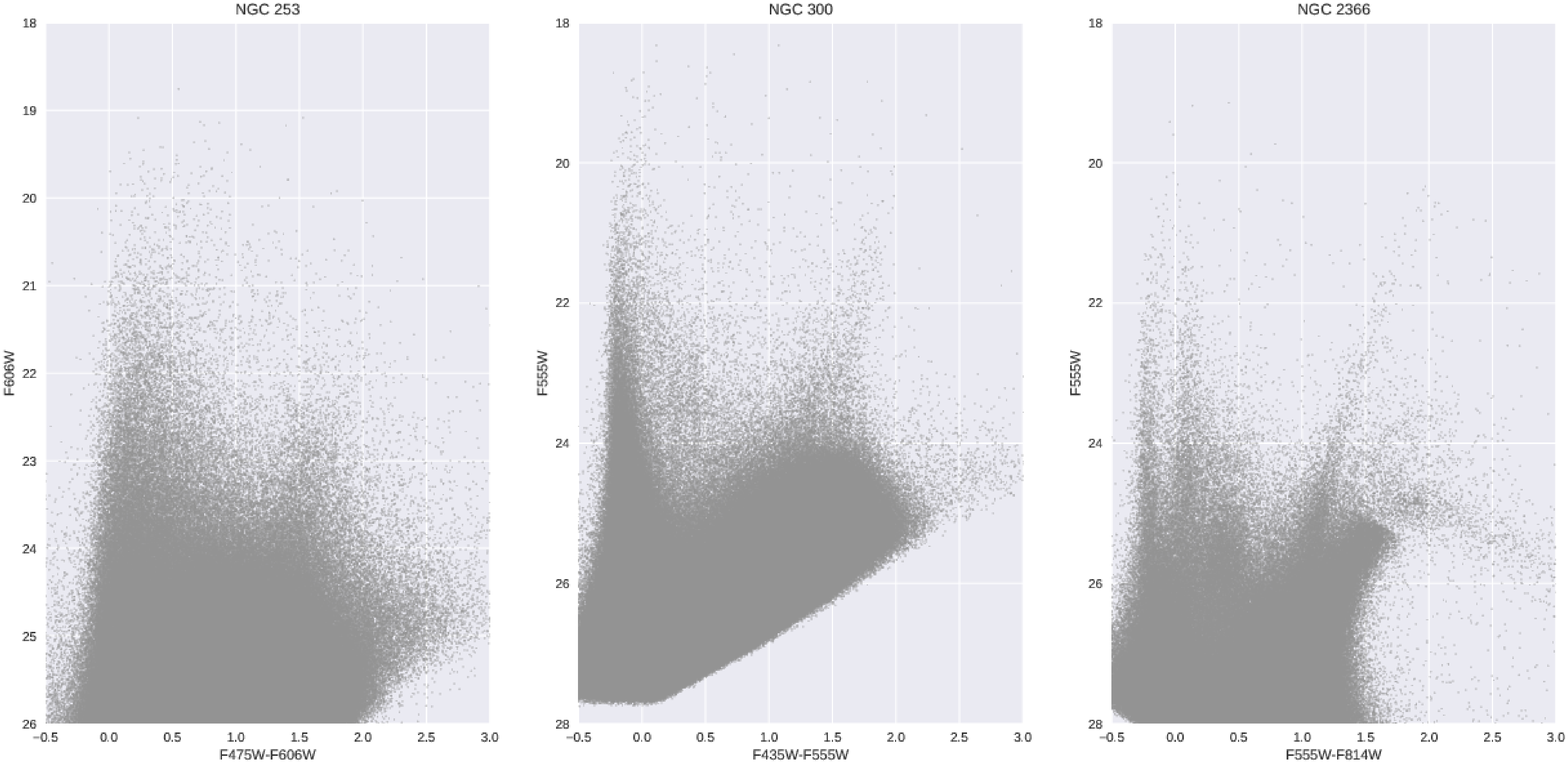}{ex_fig1}{Color - Magnitude Diagrams. Note the young main secuence that indicates a young new population of stars recently formed}

\section{Self blends on the young blue population}
In the core of a young cluster or association, where the density of blue stars is high, could the stars blend and made photometry programs to measure several false stars that are brighter than the real ones? This could be a possibility, but in this case, in the top of the main sequence (MS) stars would show in comparison with ZAMS models to be too massive or would give a wrong estimation of distance. As we considered less massive and faint stars, a red fake component would be added and it could be confused with interstellar extinction.
Again, we have not measures in any of these galaxies densities that could made a real probability to have many blends (Table 1).
But, stars that have a binary component are probable causing this kind of blends.

\section{Color Contamination}

Fig.2 are the CMDs but showing the density of stars in the color-magnitude field normalize to the total number of stars. These plots shows that the majority of the observed stars are at mag. 27 or fainter. And on the other hand very few stars are on the MS track. As the stars are so faint, a blend with a MS massive star would not change its color or magnitude in an appreciable way. Three or four magnitudes of difference is least one order of magnitude in energy flux.

	\articlefigure{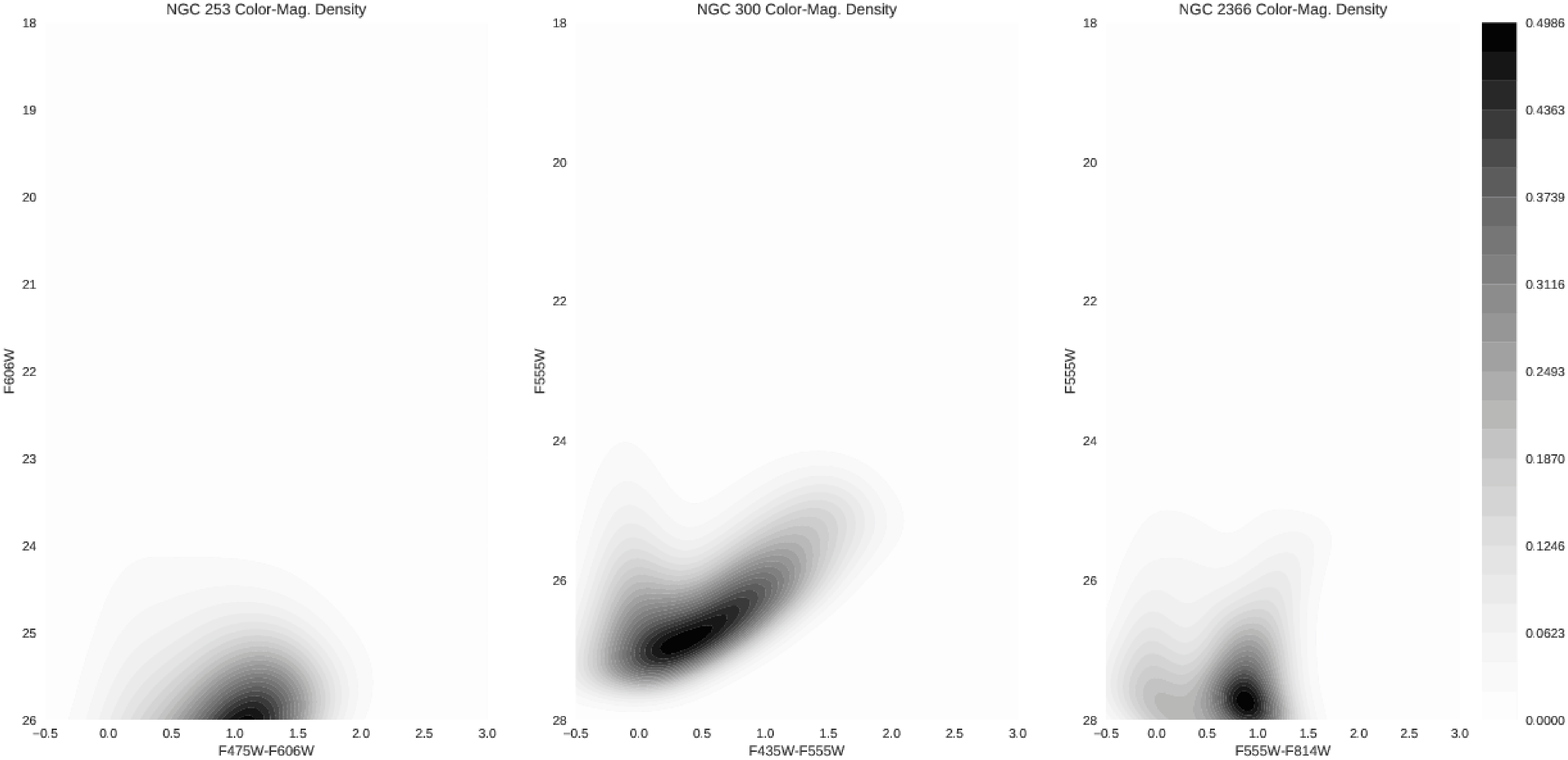}{ex_fig2}{These are the Color - Magnitude Density Diagrams. Which is basically the same data as Fig. 1 but now normalize density of stars is show. Note the regions with high density are far away from the blue massive stars main secuence.}

\begin{table}[ht]
\begin{tabular}{|l|llllll|}
\hline
Galaxy   &  Average & Max. & Filter & Limit &  Distance & Max. Prob.  \\ 
&  Density & Density & & Mag. & Mpc  &  of Blends \\
\hline
NGC 247  & 0.078  & 1.42     & F606W  & 24  & 2.55  & 6.25\ $10^{-6}$ \\ \hline
NGC 247  & 1.663  & 12.67    & F606W  & 26  & 2.55  &4.911\ $10^{-6}$ \\ \hline
NGC 253  & 0.248  & 7.61     & F606W  & 24  & 3.43  &1.788\ $10^{-4}$ \\ \hline
NGC 300  & 0.407  & 7.41     & F555W  & 25  & 1.49  &1.698\ $10^{-4}$  \\ \hline
NGC 300  & 1.617  & 13.53    & F555W  & 26  & 1.49  &5.593\ $10^{-4}$  \\ \hline
NGC 2366 & 0.064  & 0.76     & F555W  & 24  & 3.93  &1.812\ $10^{-6}$ \\ \hline
NGC 2366 & 0.581  & 3.60     & F555W  & 26  & 3.93  &4.016\ $10^{-5}$ \\ \hline
\end{tabular}
\caption{Density of stars in the sample of galaxies considering in some cases a different limit magnitude. A limiting faint magnitude would let to find more stars, but these stars are to faint that any blends with then would not produce any observable effect. Density is measure in stars arcsec$^{-2}$  }
\end{table}
	
\section{Blends and Distance}
It is easy to compute the inverse problem. This is knowing the star density, to find at what distance the observations of individual stars would be blended. Considering the real case, where ACS/HST pixels are 0".05 wide, and the density of NGC 300 (for a magnitude limit of detection of $F555W<24$). Fig. 3 shows close 9 Mpc, we are going to have a 20 \% of blend contamination and this situation would be critical in the analysis of the data. 		
\articlefigure[width=0.6\textwidth]{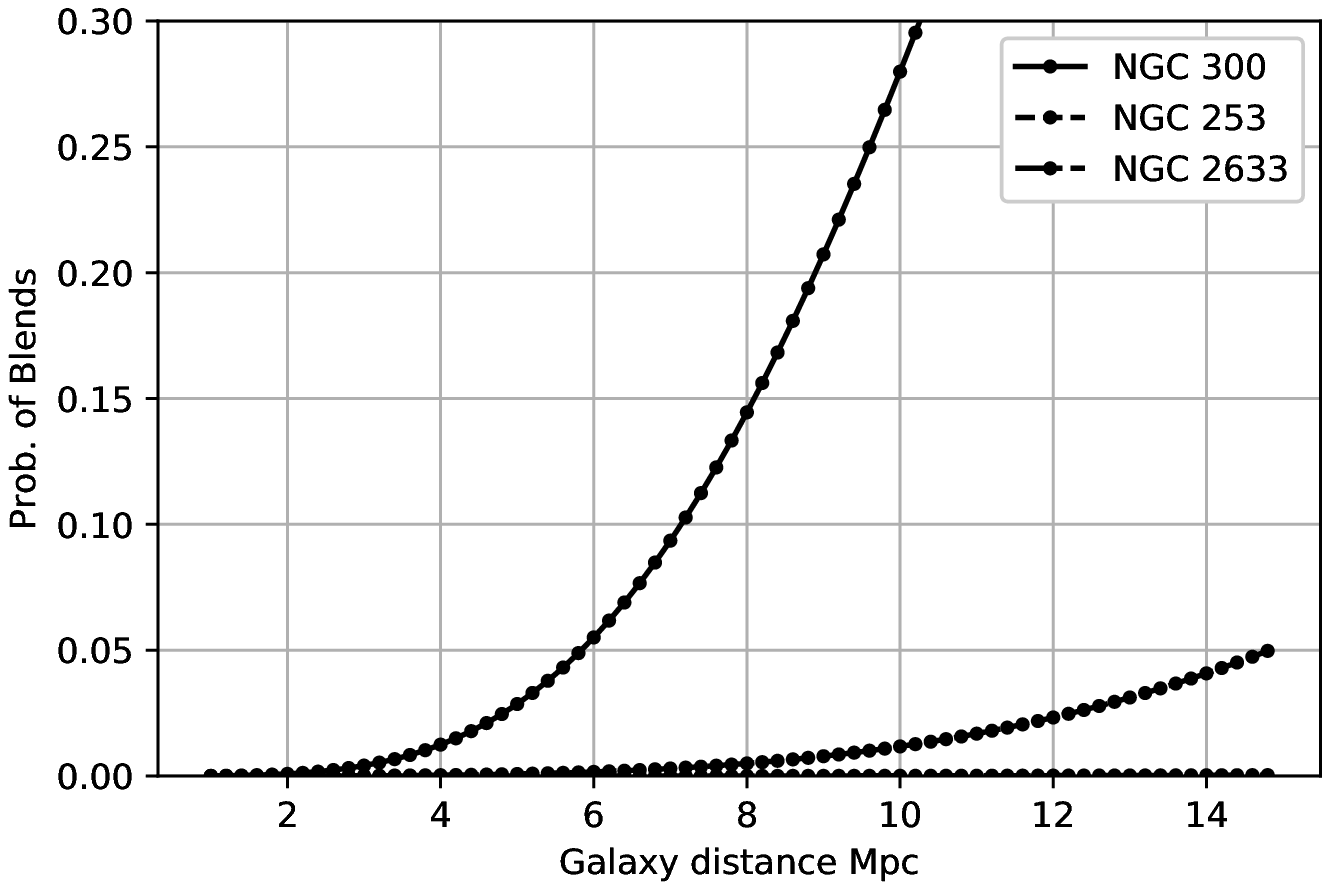}{ex_fig3}{Prob. of a blend as the distance increases}

\section{Conclusions}

\begin{itemize}

 \item For the observed density on several galaxies, we found that the undetected blends are negligible in the nearby galaxies of our sample.

\item For undetected blends to be important (for an example, like 20\% of contamination of the sample) the stellar density has to reach 0.0104 stars$\  arcsec^{-2} Mpc^{-2}$ for most dense region in a galaxy. Considering that the pixel of the ACS/HST camera is $0".05$ wide this would happen for a galaxy close to 9 Mpc in the case of NGC 300. 
 
 \item The main source of undetected blends (which is not considered in this work) are the binaries systems. Binaries need time resolved spectroscopy to be detected and are an old problem not in extragalactic data but also on galactic observations.
 
\end{itemize}	

\bibliography{P12-4}

\begin{thebibliography}{}
\expandafter\ifx\csname natexlab\endcsname\relax\def\natexlab#1{#1}\fi
\expandafter\ifx\csname url\endcsname\relax
  \def\url#1{\texttt{#1}}\fi
\expandafter\ifx\csname urlprefix\endcsname\relax\def\urlprefix{URL }\fi
\providecommand{\eprint}[2][]{\url{#2}}

\bibitem[{{Dalcanton} et~al.(2009){Dalcanton}, {Williams}, {Seth}, {Dolphin},
  {Holtzman}, {Rosema}, {Skillman}, {Cole}, {Girardi}, {Gogarten},
  {Karachentsev}, {Olsen}, {Weisz}, {Christensen}, {Freeman}, {Gilbert},
  {Gallart}, {Harris}, {Hodge}, {de Jong}, {Karachentseva}, {Mateo}, {Stetson},
  {Tavarez}, {Zaritsky}, {Governato}, \& {Quinn}}]{2009ApJS..183...67D}
{Dalcanton}, J.~J., {Williams}, B.~F., {Seth}, A.~C., {Dolphin}, A.,
  {Holtzman}, J., {Rosema}, K., {Skillman}, E.~D., {Cole}, A., {Girardi}, L.,
  {Gogarten}, S.~M., {Karachentsev}, I.~D., {Olsen}, K., {Weisz}, D.,
  {Christensen}, C., {Freeman}, K., {Gilbert}, K., {Gallart}, C., {Harris}, J.,
  {Hodge}, P., {de Jong}, R.~S., {Karachentseva}, V., {Mateo}, M., {Stetson},
  P.~B., {Tavarez}, M., {Zaritsky}, D., {Governato}, F., \& {Quinn}, T. 2009,
  \apjs, 183, 67. \eprint{0905.3737}

\bibitem[{{Kiss} \& {Bedding}(2005)}]{Kiss}
{Kiss}, L.~L., \& {Bedding}, T.~R. 2005, \mnras, 358, 883.
  \eprint{astro-ph/0501498}

\end{thebibliography}

\end{document}